\def\beq{\begin{equation}}
\def\eeq{\end{equation}}
\def\beqa{\begin{eqnarray}}
\def\eeqa{\end{eqnarray}}

\documentclass[preprint,prd,groupedaddress,showpacs,rotate]{revtex4}

\usepackage{graphicx}
\usepackage{dcolumn}

\begin{document}

\title{Deep inelastic scattering off a plasma with flavour from D3-D7 brane model}
\author{C.  A.  Ballon Bayona}
\email{ballon@cbpf.br}
\affiliation{Centro Brasileiro de Pesquisas F\'{i}sicas, Rua Dr. Xavier Sigaud 150, 
Urca, 22290-180, Rio de Janeiro, RJ -- Brazil}
\author{Henrique Boschi-Filho }
\email{boschi@if.ufrj.br} 
\affiliation{Instituto de F\'{\i}sica,
Universidade Federal do Rio de Janeiro, Caixa Postal 68528, RJ
21941-972 -- Brazil}
\author{Nelson R. F. Braga}
\email{braga@if.ufrj.br} 
\affiliation{Instituto de F\'{\i}sica,
Universidade Federal do Rio de Janeiro, Caixa Postal 68528, RJ
21941-972 -- Brazil}

\begin{abstract} 
We investigate the propagation of a space-like flavour current in a strongly coupled 
$ {\cal N}= 2 $ super Yang-Mills plasma using the D3-D7 brane model at finite temperature. 
The partonic contribution to the plasma structure functions is obtained from the 
imaginary part of the retarded current-current commutator. 
At high temperatures we find a non-vanishing result, for a high energy current, 
indicating absorption of the flavour current by the quark constituents of the plasma. 
At low temperatures there is no quark contribution to the plasma structure functions. 
\end{abstract}



\maketitle

\section{ Introduction }

Originally it was thought that the product of a heavy ion collision, the so called quark gluon plasma,
would be a weakly interacting system, behaving  approximately like an ideal gas. 
However, the experimental results from RHIC indicated that the hadronic matter formed after the collisions is strongly interacting, more similar to a perfect liquid \cite{Shuryak:2003xe}.
One expects an improvement in the understanding of the quark gluon plasma properties 
when heavy ion collisions are further investigated in the LHC. 

Strong interactions are described by QCD. However, in a strong coupling regime, as the one found in the quark gluon plasma,  its is necessary to use non-perturbative tools. A very interesting approach 
relies on the AdS/CFT correspondence which relates string theory to superconformal gauge theories at large 't Hooft coupling. 
A particular case of this correspondence is given by string theory in $AdS_5\times S^5 $ which is dual to  $ {\cal N}= 4\,\, $ $SU(N_c)$ super Yang Mills theory in four dimensions \cite{Maldacena:1997re,Gubser:1998bc,Witten:1998qj}. 
This exact correspondence inspired many interesting phenomenological models known as 
AdS/QCD that approximately describe important aspects of hadronic physics.
One of the simplest models, now called hard wall model, consists in breaking conformal invariance by introducing a hard cut off 
in the AdS space. The position of this cut off is related to an infrared mass scale in the dual  gauge theory.
The hard wall model was very successful in reproducing the scaling of hadronic scattering amplitudes at fixed angles \cite{Polchinski:2001tt,BoschiFilho:2002zs} and estimating hadronic masses  
\cite{BoschiFilho:2002ta,BoschiFilho:2002vd,deTeramond:2005su,Erlich:2005qh,DaRold:2005zs,BoschiFilho:2005yh}. 
 
A more sophisticated proposal is the D3-D7 brane model that consists in the inclusion of $N_f$ flavoured 
D7 probe branes in the $AdS_5\times S^5 $ space \cite{Karch:2002sh}. The AdS/CFT correspondence contains 
fields in the adjoint representation of the $SU(N_c)$ gauge group, like gluons, related to open strings 
attached to the D3 branes.
Including D7 branes one also has fields in the fundamental representation, like quarks,
related to open strings with an endpoint  on a D3 and the other on a D7 brane.
On the other side, mesons are described by strings with both endpoints on D7 branes. 
Masses for mesons in this model were calculated in \cite{Kruczenski:2003be}. 
For a review see \cite{Erdmenger:2007cm}.

The internal structure of hadrons can be probed by interaction processes. 
One is the deep inelastic scattering (DIS), that  was investigated using gauge/string duality in \cite{Polchinski:2002jw,BallonBayona:2007qr,Hatta:2007he,BallonBayona:2007rs,Cornalba:2008sp,Pire:2008zf, Albacete:2008ze,Gao:2009ze,Hatta:2009ra,BallonBayona:2008zi,Cornalba:2009ax}. 
Elastic form factors also give information about the hadronic structure. Form factors in AdS/QCD  models 
were studied for example in \cite{Hong:2003jm,Grigoryan:2007vg,Grigoryan:2007my,Brodsky:2007hb,Kwee:2007nq,RodriguezGomez:2008zp,Bayona:2009ar}. 
Other aspects of hadronic interaction processes using  AdS/QCD have been discussed in, for example  
\cite{Hatta:2008tx,Hatta:2008tn,Hatta:2008qx,Vega:2009zb}.

The AdS/CFT correspondence can be extended to describe gauge theories at finite temperature introducing  a black hole in the AdS geometry \cite{Witten:1998zw,Son:2002sd,Kovtun:2005ev}. This version of the correspondence describes the strongly coupled $ {\cal N} = 4$ super Yang Mills plasma, since the particles 
are in a deconfined phase. 
The structure of this plasma was investigated in \cite{Hatta:2007cs} considering a DIS process of 
an ${\cal R}$  current. The gravity dual of this current is a gauge field propagating in the black hole AdS
space. 
It was found that at high energies, the  ${\cal R}$  current probes the partonic behaviour of the plasma
(associated with gluons),  giving non-vanishing structure functions. 
The fact that this result occurs at any non-zero temperature is  consistent 
with the non-confining character of the AdS-black hole model.  

In this article we are going to study the structure of a strongly coupled plasma containing flavour degrees of freedom 
using the D3-D7 brane model at finite temperature. This consists in the inclusion of $N_f$ coincident D7 probe branes in the black hole  AdS space \cite{Babington:2003vm,Mateos:2006nu,Hoyos:2006gb,Mateos:2007vn}.
We consider the propagation of a space-like gauge field living in the D7 branes, corresponding to the DIS of a flavour current in a strongly coupled   $ {\cal N}= 2 $ super Yang-Mills plasma.  
 There are two different thermal phases in the D3-D7 model: the low temperature Minkowski embedding when the D7 branes do not touch the horizon and the high temperature black hole embedding when the D7 branes have an induced horizon. 
We calculate in both phases the quark contribution to the plasma structure functions, considering the absorption of a flavour current by the quark constituents of the plasma. 
We find non-vanishing results in the high temperature phase for a high energy current.  
In the low temperature regime, corresponding to the Minkowski embedding, we find that the flavour current is not absorbed by the quark constituents of the plasma.

Note that the D3-D7 model contains both the $\cal R$ and flavour currents. 
The  ${\cal R} $ current is dual to a gauge field propagating in the bulk geometry generated by the D3 branes  
while the flavour current dual is a gauge field propagating on the world volume of the D7 branes. 
This way, studying a DIS process in this model we can unveil not only the partonic 
structure associated with gluons but also the partonic structure associated with quarks.

This article is organized as follows. In section {\bf II}, we briefly review the ${\cal R}$ current results. 
Then, we study, in section {\bf III}, the equations of motion of a gauge field on the D7 branes. 
In section {\bf IV} we analyse the effective potentials at low and high temperatures and 
calculate the quark contribution to the plasma structure functions in these regimes. 
We end, in section {\bf V}, with our conclusions.


\section{ Deep inelastic scattering of an ${\cal R}$ current }

In this section we briefly review the calculation
of the retarded current-current correlator of an ${\cal R} $ current and the corresponding DIS structure functions. According to the AdS/CFT correspondence at finite temperature, an ${\cal N}  = 4 $ super Yang Mills 
$SU(N_c)$ plasma is dual to a  black hole in $AdS_5\times S^5$ space. This space can be described by the metric  
\begin{equation}
\label{background} ds^2 \,= \,\frac{r_0^2}{L^2 u} 
\Big[ - (1-u^2) dt^2 \,+\,  d{\bar x}^2 \Big] 
\,+\,   \frac{L^2}{4u^2 (1 - u^2)} du^2\,+\, L^2 d \Omega^2_5 \,,
\end{equation}

\noindent where $ r_0 \,=\, \pi L^2 T $ , \, $L $ is the AdS radius ($L^4 \,=\, 4\pi \alpha'^2 g_{YM}^2 N_c $), \,   $T$ is the temperature and $ d \Omega^2_5 $ is the $S^5 $ metric. The radial coordinate $u$ is dimensionless, with the horizon located at $u = 1 $ and the boundary at $u =0 $. 

The vector field $ A_m = ( A_u , A_\mu ) $ dual to the ${\cal R}$-current is described by the supergravity
action  
\begin{equation}
\label{Action5d}
S \,=\,-\, \frac{ N_c^2}{ 64 \pi^2 L } \int d^{4}x du \sqrt{-g}\,g^{mr} g^{ns}   \, F_{mn} F_{rs} \,\,.
\end{equation}

\noindent Choosing the gauge $A_u = 0$ and the plane wave solution 
\begin{equation}
A_\mu ( x ,  u) \,=\, e^{-iq \cdot x  } {\bar A}_\mu (u) \,\,,
\label{planewave}
\end{equation}

\noindent with $q^\mu = (w, 0,0,q) $, the on shell action takes the form  
\begin{equation}
\label{OnshellAction}
S \,=\,-\, \frac{ N_c^2 T^2}{ 16 } \int d^{4}x  \Big[ \Big( A_t + \frac{ {\bar w}}{\bar q} A_3 \Big) 
\, \partial_u A_t \,-\, A_i \partial_u A_i  \Big]_{u = 0 } \,\,, 
\end{equation}

\noindent where  $i =1,2$, $ {\bar w} \equiv \frac{w}{2\pi T} \,$ , $\,\,  {\bar q}  \equiv \frac {q} { 2\pi T }\,\,$  and we used the relation 
$ \partial_u A_t = ({\bar q }/{\bar w}) \partial_u A_3 $ valid at the boundary $u =0$.

The retarded current-current commutator of the boundary gauge theory is defined by
\begin{equation}
\label{Rcommutator}  R_{\mu\nu} (q) \,=\, i \int d^4x  \,e^{-i q\cdot x } \, \theta ( t ) 
\langle [ J_\mu (x) , J_\nu (0) ] \rangle 
\end{equation}

\noindent and can be decomposed as  
\begin{equation}
R_{\mu\nu} \, = \,  \Big( \eta_{\mu\nu} \,-\, \frac{q_\mu q_\nu}{Q^2} \, \Big) R_1
\,+\, \Big[ n_\mu n_\nu \,- \, \frac{ n \cdot q }{Q^2} ( n_\mu q_\nu + n_\nu q_\mu ) \,+ 
\frac{q_\mu q_\nu}{(Q^2)^2} ( n\cdot q)^2 \, \Big] R_2\, ,
  \label{decomp}
\end{equation}

\noindent where $n^\mu $ is the four velocity of the plasma and $Q^2 $ is the virtuality 
defined by $ Q^2 \equiv q^2 - w^2 $. In the plasma rest frame $n^\mu=(1,0,0,0)$. 
The DIS structure functions of the plasma are given by
\begin{equation}
F_1 = \frac{1}{2\pi} {\rm Im} \, R_1 \hskip 1cm F_2 = -  \frac{ n\cdot q }{2\pi T } \, {\rm Im} \, R_2 \,.
\label{defF1F2}
\end{equation}
These structure functions can be obtained from those of the DIS off a hadron identifying
the hadronic momentum $P^\mu$ with the plasma momentum $Tn^\mu$.

Following a supergravity prescription similar to that of ref. \cite{Son:2002sd}, the retarded current-current commutator is obtained by differentiating the boundary action density 
with respect to the boundary values of the vector fields
\begin{equation}
\label{Rcommutator2}  R_{\mu\nu} (q) \,=\, \frac{\partial^2 {\cal S }}{ \partial A^{(0)}_\mu \partial A^{(0)}_\nu} \;; \qquad \qquad
(A^{(0)}_\mu =\lim_{u\to 0} A_\mu)\;,
\end{equation}
after imposing an {\it ingoing} condition for the gauge fields at the horizon, meaning that there is no  reflection. 

The equations of motion for the components of the gauge fields can be written as Schr\"odinger equations.
For the longitudinal part it reads
\begin{equation}
\label{eqlong}
\frac{ \partial^2 \psi}{\partial u^2} - V (u)  \psi \,=\,0\,,
\end{equation}

\noindent where $ \psi (u) = \sqrt{ u (1-u^2)}\, \partial_u A_t  \,$  and the effective potential is
\begin{equation}
\label{Rlogpotential}
V(u) =  \frac{1}{ u (1 - u^2)^2 }\,\Big[ -  \frac{1}{4u} (1 + 6u^2 - 3u^4) + {\bar Q}^2 - {\bar q}^2 u^2 \Big]\,,
\end{equation}

\noindent with ${\bar Q}^2 \equiv {\bar q}^2 - {\bar w}^2 $.
For the transversal components we have
\begin{equation}
\label{eqtransv}
\frac{ \partial^2 \phi}{\partial u^2}  - {\tilde V} (u)  \phi \,=\,0\,,
\end{equation}

\noindent where $ \phi (u) = \sqrt{(1-u^2)}\,A_i (u)  \,$ and   
\begin{equation}
\label{Rtranspotential}
{\tilde V} (u) \,=\, \frac{1}{ u (1 - u^2)^2 }\,\Big[ {\bar Q}^2 - {\bar q}^2 u^2 - u  \Big] 
\end{equation}

Analysing the behaviour of the potentials (\ref{Rlogpotential}) and (\ref{Rtranspotential}) 
one finds different regimes depending on the relation between ${\bar q}$ and ${\bar Q}\,$ \cite{Hatta:2007cs}. These regimes are separated by the critical value 
$ {\bar q}/{\bar Q}^3 = 8/(3\sqrt{3}) \approx 1.54 $. At low energies, $ {\bar q}/{\bar Q}^3 \ll 1 $,  there is no relevant contribution to the structure functions since the imaginary part of $R_{\mu\nu}$ is negligible.
At high energies  $ {\bar q}/{\bar Q}^3 \gg  1 $, a significant imaginary term arises in $R_{\mu\nu}$ 
leading to the non vanishing  structure functions
\beqa
F_1 & \approx &  \frac{3 }{16 \Gamma^2(1/3)} \,  N_c^2  T^2 \left( \frac{ Q^2}{12 \pi T^2 x} \right)^{2/3} \nonumber \\
F_2 & \approx &   2 x F_1 \,,
\eeqa

\noindent   where $x$ is the Bjorken variable that in the plasma rest frame takes the form $ x = Q^2 / 2 w T\,$.  These structure functions imply dissipation of the ${\cal R}$  current,
unveiling the existence of a partonic structure associated with gluons in the ${\cal N}  = 4 $ super Yang Mills
plasma, as discussed in ref. \cite{Hatta:2007cs}. Note that these results for the ${\cal R}$  current are also valid in the ${\cal N}  = 2 $ super Yang Mills plasma of the D3-D7 brane model.


\section{ Deep inelastic scattering of a flavour current }

\subsection{ The D7 brane embedding } 

We place $N_f$ coincident D7  probe branes in the black hole $AdS_5\times S^5$ space of eq.  (\ref{background}). 
In order to describe the D7  probe branes, we decompose the metric of the sphere $S^5 $  as
\begin{equation}
\label{S5} d \Omega^2_5 \,= \,  d\theta^2 
\,+\, \sin^2\theta d \Omega^2_3 
\,+\, \cos^2\theta d\varphi^2  \,.
\end{equation}
 
The usual choice is to fix $\varphi=0$ and $\theta=\theta(u)$. 
Then, each D7 brane is described by the metric   
\begin{equation}
\label{D7metricu} ds^2 \,= \,\frac{r_0^2}{L^2 u} 
\Big[ - (1-u^2) dt^2 \,+\,  d{\bar x}^2 \Big] 
\,+\, L^2 \Big[ \frac{1}{4u^2 (1 - u^2)} \,+\, 
 \theta'^2  \Big]du^2\,+\, L^2\sin^2\theta d \Omega^2_3 \,.
\end{equation}
Here, and from now on, prime denotes differentiation with respect to the variable $u$. 
The location of the D7 branes is contained in $\theta=\theta(u)$. 
This function is obtained by solving the equation of motion that comes from the brane action:
\beqa
S_{D7} &\,=\,& \,-\, N_f \, \mu_7  \int d^8 x  \sqrt{- \det g} \nonumber \\
&\,=\,& \,-\, N_f \, \mu_7 \pi^2 r_0^4   \int d^4x \,  
\int_0^1 \frac{du }{u^3 } \sin^3 \theta  \sqrt{ 1 + 4u^2 (1 - u^2 ) \theta'^2 } \,.
\eeqa
where $\mu_7\,=\, [(2 \pi)^7 g_{YM}^2\, \alpha'^4 ]^{-1}$ is the tension of each D7 brane. 
The D7 brane equation of motion reads 
\begin{equation}
\label{eqtheta}
3 \cos \theta \Big[ 1 + 4u^2 (1 - u^2 ) 
\theta'^2 \Big]\,+\, 4  u \sin \theta \Big[ -u (1 -u^2 ) \theta'' + 4 u^2 ( 1 -u^2 ) ( 2 - u^2 ) \theta'^3 
+ ( 1 + u^2 ) \theta' \Big] = 0\,. 
\end{equation}

There are solutions of three different kinds.  In the first  one,  
the branes touch the black hole horizon. 
This solution is called black hole embedding. 
An observer living on the branes world volume will see an induced horizon. 
In the  second kind of solutions the branes never reach the horizon. This solution is known 
as Minkowski embedding. In this case, an observer on the branes do not see any induced horizon. 
The third solution is a critical embedding that connects the black hole and Minkowski embeddings.

Near the boundary all the solutions have the asymptotic expansion
\begin{equation}
\theta (u)\, = \, \frac{\pi}{2} \,-\, \frac{m}{\sqrt{2}} u^{1/2} \,-\, 
\frac{1}{2\sqrt{2}} ( \frac{ m^3}{6} + c \,) u^{3/2} + ...\,\, ,
\label{theta}
\end{equation}

\noindent  where $ m \equiv  M /T $  with $ M $  proportional to the mass gap at zero temperature and $c$ is the quark condensate. Below we briefly review the differences between these three embeddings in the region far from the boundary and the phase transition between the Minkowski and black hole embeddings, found in \cite{Mateos:2007vn}.

\subsubsection{The black hole embedding}

The black hole embedding is characterized by the condition that the coordinate $\theta (u)  $ goes to a constant value $ \theta_0 $ different from zero  at the horizon: \, $\theta (1) = \theta_0$ .

It is convenient to use the variable $\chi (u)  \equiv \cos \theta (u) $ which has the near horizon expansion 
 \beq
\chi (u)  = \chi_0 + \chi_1 (1-u) + \chi_2 (1-u)^2 + \dots  \, ,
\eeq
with $0 \le \chi_0 < 1$.  Using this expansion in (\ref{eqtheta}) we obtain 
\beq
\chi_1 = - \frac38 \chi_0 \quad \rightarrow \quad \theta'(1) = - \frac38 \cot \theta_0 \,.
\eeq 

We use the above conditions for $\theta$ and $\theta'$  as initial conditions at $u=1$  
for the numerical integration of eq. (\ref{eqtheta}). 

\begin{figure}[ht]
\includegraphics[width=60mm]{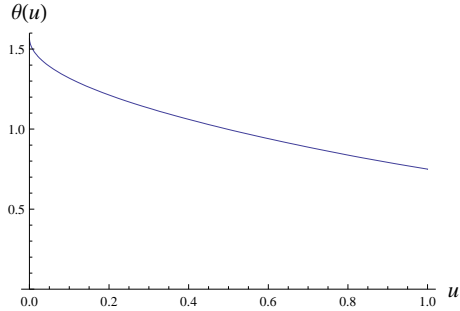}
  \caption[]{Black hole D7 metric for $\theta_0\,=\,0.75$\,.}
  \label{BHD7}
\end{figure}

A typical solution is illustrated on Fig. \ref{BHD7}. 
As shown in \cite{Mateos:2007vn}, black hole embeddings exist only for temperatures higher than $ 0.764 M$.

\subsubsection{The Minkowski embedding}

On the other hand,  in the case of Minkowski embedding the coordinate $u$ has a maximum value $ u_0 < 1$ where
the coordinate $\theta$ vanishes: $ \theta (u_0 ) = 0  $. The region $ u = u_0 $ is interpreted as the tip of the branes.  
The requirement of smoothness without conical deficit at the tip of the branes leads to  $ \theta' (u_0) = - \infty $ \cite{Hoyos:2006gb}. 

These conditions on $\theta$ and $\theta'$ suggest the near horizon expansion 
\beq
\theta = \left (1-\frac{u}{u_0} \right )^{\alpha} \left[ \rho_0 + \rho_1 \left(1 - \frac{u}{u_0}\right) + \rho_2 \left ( 1 -  \frac{u}{u_0} \right )^2 + \dots \right ] \, , 
\eeq
with $0 < \alpha < 1$.Using this expansion in (\ref{eqtheta}) we obtain 
\beq
\alpha = \frac12 \quad , \quad \rho_0 = \sqrt{ \frac{2}{2- u^2_0}} \,. 
\eeq

\begin{figure}[ht]
\includegraphics[width=60mm]{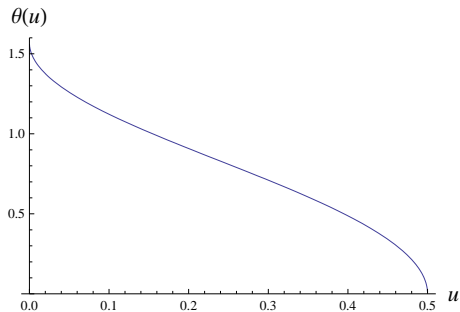}
  \caption[]{Minkowski D7 metric for $u_0\,=\,0.5$.}
 \label{MinkD7}
\end{figure}

In the numerical integration we evaluate this asymptotic expression and its derivative at some $u$ very close (but not equal) to $u_0$. A typical solution is shown in Fig. \ref{MinkD7}. 

As shown in \cite{Mateos:2007vn}, Minkowski embeddings exist only for temperatures lower than $ 0.772  M $.

\subsubsection{The critical embedding} 

\begin{figure}[ht]
\includegraphics[width=60mm]{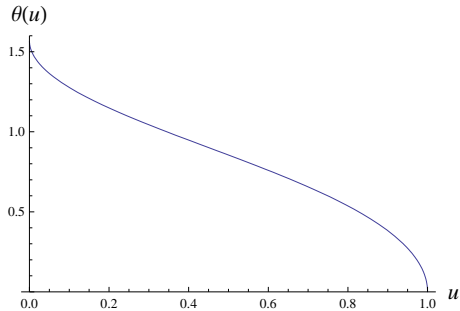}
  \caption[]{Critical D7 metric corresponding to $\theta_0\,=\,0$ or $u_0\,=\,1$ .}
 \label{CriticalD7}
\end{figure}

The critical embedding appears as an intermediate solution between black hole embeddings and Minkowski embeddings.
In this case the coordinate $u$ reaches the horizon when the coordinate $\theta $ goes to zero.
This is shown in Fig. \ref{CriticalD7}.

\subsubsection{The phase transition} 

The Minkowski and black hole embeddings can coexist for temperatures between $ 0.764 M $ and $ 0.772  M$. However, the physical embedding is the one that minimizes the free energy at  a given temperature. It was shown in \cite{Mateos:2007vn} that a first order  transition occurs at a temperature $T_c = 0.766 M $. Below $T_c $ the Minkowski embedding has the lower free energy so it is thermodynamically preferred while the black hole embedding is dominant above $T_c$. 

A consequence of this first order phase transition is a finite jump in the quark condensate at $T=T_c$, where the D7 branes  go from a Minkowski embedding to a black hole embedding. We will see below how the abrupt change in the D7 brane embedding has a non-trivial consequence on the partonic contribution to the structure functions of 
the ${\cal N}= 2 $ super Yang-Mills plasma when probed by a flavour current.

\subsection{Gauge field equations and Schr\"odinger potentials}

We consider a flavour current $ U(1) \subset U(N_f ) $. The gravity dual of this current 
is a gauge field fluctuation living in the world volume of the D7 branes. From the Dirac Born Infeld action 
one finds the gauge field kinetic term 
\begin{equation}
\label{GFaction}
S \,=\,-\, N_f  \mu_7 \, \pi^2 \alpha'^2   \int d^{8}x \sqrt{-G}\,  \, F_{MN} F^{MN} \,\,, 
\end{equation}

\noindent where $  F_{MN} = \partial_M A_N - \partial_N A_M \,$ and 
$A_M = (A_\mu, A_u , A_\alpha )\,$.  
We choose $A_{\alpha}=0$, $\partial_{\alpha}A_\mu = 0$, $ A_u = 0 $ and the 
plane wave ansatz
\begin{equation}
A_\mu (x,u)\,=\, \exp\{ -iwt + i qx^3 \} \,{\bar A}_\mu (u) \,.
\end{equation}

\noindent The equations of motion take the form
\begin{eqnarray}
\partial_u A_3 &=& -\, \frac{\bar w}{{\bar q} (1- u^2)}  \partial_u A_t \label{longeq1}
\\
\frac{\Gamma^{1/2}}{\sin^3 \theta } \partial_u  \Big[ \frac{\sin^3 \theta }{\Gamma^{1/2}} 
\partial_u A_t \Big] &-& \frac{\Gamma}{u (1 - u^2 ) } \Big[ {\bar q}^2 A_t + {\bar q} {\bar  w} A_3 \Big] 
\,=\,0 \label{longeq2}
\\
\frac{\Gamma^{1/2}}{\sin^3 \theta (1 - u^2 ) }  \partial_u \Big[ 
\frac{\sin^3 \theta (1 - u^2 ) }{\Gamma^{1/2}} \partial_u A_i \Big]
&-& \frac{\Gamma}{u (1-u^2)^2 } \Big[ - {\bar w}^2 \,+\,  (1-u^2) {\bar q} ^2  \Big] A_i  \,=\,0\;, \;\;\; \cr 
& & \label{transverseeq}
\end{eqnarray} 
where $ \Gamma \equiv 1 + 4 u^2 (1 - u^2 ) \theta'^2$ \,.

In order to gain intuition on the problem, we write the gauge field equations (\ref{longeq1}), (\ref{longeq2}) and (\ref{transverseeq}) into a Schr\"odinger form, as in ref. \cite{Myers:2007we}. 
We will find two relevant potentials whose form determine the behavior of the retarded current-current commutator and hence the partonic contribution to the DIS structure functions. 


\subsubsection{Longitudinal component }

Using $\, a(x,u) \equiv \partial_u A_t \,$, we find from eqs. (\ref{longeq1}) and (\ref{longeq2}) the second order 
differential equation
\begin{equation}
\partial^2_u a \,+\, \frac{1}{H_1} \partial_u H_1 \partial_u a \,-\, \frac{1}{H_0} \Big[ H_3 + 
H_2 \, {\bar q}^2 \,-\, {\bar w}^2 \, \Big] a \,=\,0 \,, \label{longitudinalfieldeq}
\end{equation}
where 
\begin{eqnarray}
H_0 &=&  \frac{u(1 - u^2)^2 }{\Gamma}   \nonumber\\
H_1 &=& \frac{ u (1-u^2) \sin^3 \theta}{ \Gamma^{3/2}}
\nonumber\\  H_2 &=&  1 - u^2 \nonumber\\
H_3 &=& - (1 - u^2) \partial_u \Big[ \frac{u - u^3}{\Gamma} \Big( 3 \,\cot\theta \,\,\theta^{\prime}
\,-\,\frac{\Gamma^{\prime}}{2 \Gamma}  \Big) \,\Big]\;.
\end{eqnarray}

Introducing: $a(x,u) = {\eta} (u) \psi (x,u) $ and imposing that
\begin{equation}
 \frac{\eta'}{\eta} = - \frac 12 \frac {H_1'}{H_1} \,, \label{etaeq}
\end{equation}
we find a Schr\"odinger equation 
\begin{equation}\label{Schroed}
\frac{ d^2 \psi }{du^2} \,-\,V (u)  \,\psi \,=\,0 \,,
\end{equation}
with potential 
\begin{eqnarray}
V(u)  &=& \frac{1}{2} \Big( \frac{H_1^{\prime}}{H_1} \Big)^{\prime} 
+ \frac{1}{4} \Big( \frac{H_1^{\prime}}{H_1} \Big)^2 
+ \, \frac{1}{H_0} \Big[ H_3 + 
H_2 {\bar q}^2 \,-\, {\bar w}^2 \, \Big] \nonumber\\
 &=&  \frac{3u^4-6u^2-1}{4u^2(1-u^2)^2} \, 
+ \, \frac{\Gamma}{u(1-u^2)^2}\Big[{\bar Q}^2 - u^2{\bar q}^2\Big] \, 
- \frac 32 \cot\theta \, \theta''-\frac {\Gamma''}{4\Gamma} 
\,-\, \frac {\Gamma'}{4\Gamma}\frac{1-3u^2}{u(1-u^2)} 
\nonumber\\
&+& \frac 32 \cot\theta \, \theta^{\prime}\Big[ \frac {\Gamma'}{2\Gamma} 
\,-\,\frac{1-3u^2}{u(1-u^2)}\Big]
\,+\, \frac 32\Big( 1 \,+\, \frac {5}{2} \cot^2\theta \Big){\theta'}^2
\,+\, \frac 5{16} \Big(\frac {\Gamma'}{\Gamma}\Big)^2\,.
\end{eqnarray}
A possible solution to eq.(\ref{etaeq}) is  ${\eta} = C ( - H_1 )^{-1/2} $, with $C$ an arbitrary constant, 
chosen for convenience as $C\,=\,2 \sqrt{2} r_0^2/ L^2$. 
Near the boundary and in the Bjorken limit $\bar Q^2 \gg m^2$,  the longitudinal potential can be approximated by 
\begin{eqnarray}
V(u) &=& - \frac{1}{4u^2} \, 
+\,   \frac{{\bar Q}^2}{u}\, 
+ \frac{m^2}{2} {\bar Q}^2 \,
- {\bar q}^2 u  \,-\, \frac{m^2}{2} {\bar q}^2 u^2 \,. \label{approxlongpot}
\end{eqnarray}
The terms containing $m^2$ represent small perturbations of the near boundary approximation of 
the potential (\ref{Rlogpotential}). If we consider the interesting high energy regime $\bar q \gg \bar Q^3$ this potential reduces to  
\begin{eqnarray}
V(u) &=& - \frac{1}{4u^2} \, 
- {\bar q}^2 u  \,-\, \frac{m^2}{2} {\bar q}^2 u^2\,. \label{approxlongpothighenergy}
\end{eqnarray}

 \noindent This expression will be useful for the calculation of the DIS structure functions in section~{\bf IV}.


\subsubsection{Transversal component }

In a similar way, for the transversal modes, eq. (\ref{transverseeq}) can be rewritten as 
\begin{equation}
\partial^2_u A_i \,+\, \frac{1}{\tilde H_1} \partial_u {\tilde H}_1 \partial_u A_i \,-\, \frac{1}{H_0} \Big[ H_2 \, {\bar q}^2 \,-\, {\bar w}^2 \, \Big] A_i \,=\,0\,, \label{transversefieldeq}
\end{equation}
where 
\begin{eqnarray}
{\tilde H}_1 &=& \frac{(1-u^2) \sin^3 \theta}{\Gamma^{1/2}}
\,.
\end{eqnarray}

Defining $\, A_i(x,u)\,=\,\epsilon(u)\,\Phi_i(x,u) \,$ and imposing 
\begin{equation}
 \frac{\epsilon'}{\epsilon } = - \frac 12 \frac {{\tilde H_1'}}{{\tilde H_1}}\,, \label{eqepsilon}
\end{equation}

\noindent we find a Schr\"odinger equation for $\Phi_i(x,u)$, similar to eq. (\ref{Schroed}),  with the potential
\begin{eqnarray}
\tilde V (u) &=& \frac{1}{2} \Big( \frac{{\tilde H}_1^{\prime}}{{\tilde H}_1} \Big)^{\prime} 
+ \frac{1}{4} \Big( \frac{{\tilde H}_1^{\prime}}{{\tilde H}_1} \Big)^2 
+ \, \frac{1}{H_0} \Big[\,H_2 {\bar q}^2 \,-\, {\bar w}^2 \, \Big] 
\nonumber\\
&=& - \frac{1}{(1-u^2)^2} \, 
+ \, \frac{\Gamma}{u(1-u^2)^2}\Big[{\bar Q}^2 - u^2{\bar q}^2\Big] \, 
+ \frac 32 \cot\theta \, \theta'' -\frac {\Gamma''}{4\Gamma} 
\,+\, \frac{u}{(1-u^2)}   \frac {\Gamma'}{2\Gamma}  
\nonumber\\
&-& \frac 32 \cot\theta \, \theta^{\prime} \Big[   \frac {\Gamma'}{2\Gamma} 
\,+\,\frac{2u }{(1-u^2)}\Big]
\,+\, \frac 32\Big( - 1 \,+\, \frac {1}{2} \cot^2\theta \Big){\theta'}^2
\,+\, \frac 5{16} \Big(\frac {\Gamma'}{\Gamma}\Big)^2\,.
\end{eqnarray}

\noindent The solution to eq. (\ref{eqepsilon}) is $\epsilon\,=\, \tilde C (\tilde H_1)^{-1/2}$ and we choose for convenience $\tilde C \,=\, \sqrt{2} r_0$. Again, near the boundary and for $\bar Q^2 \gg m^2$,  we find the approximate transversal  potential 
\begin{eqnarray}
{\tilde V} (u) &=&  
   \frac{{\bar Q}^2}{u}\, 
+ \frac{m^2}{2} {\bar Q}^2 \,
- {\bar q}^2 u  \,-\, \frac{m^2}{2} {\bar q}^2 u^2\,. \label{approxtransvpot}
\end{eqnarray}
In the high energy  regime $\bar q \gg \bar Q^3$ this potential can be approximated by 
\begin{eqnarray}
{\tilde V} (u) &=&  
   - {\bar q}^2 u  \,-\, \frac{m^2}{2} {\bar q}^2 u^2\,. \label{approxtransvpothighenergy}
\end{eqnarray}


\subsection{The retarded flavour current commutator}

The on-shell action for the gauge field has the following boundary term   
\beqa
S_{B} \,=\,8 \pi^4 \alpha'^2 \, N_f \, \mu_7 \,  \,  r_0^2  \left\{ \frac{\sin^3 \theta }{\Gamma^{1/2}} 
\int d^{4}x  [\, - (A_t + \frac {\bar  w}{\bar q} A_3) \, a + (1-u^2) A_i \partial_u A_i ] 
\right\}_{u = 0} , 
\eeqa
where  $i = 1,2$. The hypersurface $u = 0 $ defines the boundary which excludes the horizon or the tip of the D7 branes. 

We will use this boundary action to calculate the quark contribution to the plasma structure functions.
The near boundary expansions for the fields $a$ and $A_i$ that satisfy the equations of motion are given by 
\beqa
a(x,u)&\,=\,& \bar q^2 \left [  A^{(0)}_t + \frac{\bar w}{ \bar q} A^{(0)}_3 \right] \, \left \{ a_0 \, + \, \ln{u} \,+\,  a_1 u   \,+\,  b_1 u \ln{u}  \,+\, {\cal O}(u^2) \,+\, {\cal O}(u^2 \ln{u})  \right \} \nonumber \\
A_i(x,u) &\,=\, & A^{(0)}_i \left \{ 1 \,+\, \bar Q^2 [  c_1 u \,+\,  d_1 u \ln{u} ] \,+\, {\cal O}(u^2) \,+\, {\cal O}(u^2 \ln{u})  \right\} \label{aAi}
\eeqa
where $A^{(0)}_\mu(x)\,= \,A_{\mu}(x,u)\vert_{_{u=0}}$. 
The coefficients $a_0,\,a_1,\, b_1,\,c_1,\, , d_1 $ can be obtained from 
the differential equations  (\ref{longitudinalfieldeq}) and (\ref{transversefieldeq}). 
With these expansions the boundary on shell action can be written as 
\beqa
S_{B}\,=\, \frac{N_f N_c}{16 \pi^2 }  \,  
\int d^{4}x  \Big \{ \, -  q^2\,[A^{(0)}_t + \frac {\bar  w}{\bar q} A^{(0)}_3]^2 \, ( \,\ln{u}+ a_0\,) \,\nonumber \\
 + \, Q^2 A^{(0)}_i  A^{(0)}_i \,(\,d_1 \ln{u} + c_1+d_1\,) \Big\}_{u=0} 
\, ,
\eeqa
The divergent logarithmic terms can be canceled by the following (boundary) counterterms 
\beqa
S_{ct}\,=\,  \frac{N_f N_c}{16 \pi^2}  \,
\int d^{4} x  \Big \{ \,  q^2\,[A^{(0)}_t + \frac {\bar  w}{\bar q} A^{(0)}_3]^2 \, \left[ \, \ln{u}+ 2 \gamma + 2 \ln{(\frac{M}{2\pi T})}  \right] \,\nonumber \\
 - \, Q^2 \, A^{(0)}_i  A^{(0)}_i \,d_1 \,   \left[  \ln{u}+2\gamma + 2 \ln{(\frac{M}{2\pi T})} \right] \Big\}_{u=0} 
\, ,
\eeqa
where we have included in the counterterms the temperature $T$ and the infrared scale $M$ is the D7 brane mass scale at zero temperature defined in the previous section.  We also included the Euler-Mascheroni constant $\gamma$ as is done in \cite{Hatta:2007cs}. Summing both terms we obtain the total finite action 
\beqa
S_{Total}&=&  \frac{N_f N_c}{16 \pi^2}  \,
\int d^{4}x  \Big \{ \,  \left(\eta_{\mu \nu} - \frac{q_\mu q_\nu}{Q^2} \right) Q^2 \,\left[\,c_1+d_1 - 2\gamma \, d_1 - 2 \ln{(\frac{M}{2\pi T})} \, d_1 \right] \nonumber \\
&+& (n_\mu-q_\mu \frac{n \cdot q}{Q^2})(n_\nu-q_\nu \frac{n \cdot q}{Q^2}) \, \frac{Q^4}{q^2} \left [ c_1+d_1-a_0 + 2 \ln{(\frac{M}{2\pi T})} (1-d_1) + 2\gamma (1-d_1)  \right ] \Big \}\nonumber\\
&\times&  A^{(0) \mu}A^{(0) \nu} 
\eeqa
So, differentiating this action with respect to the boundary values of the gauge fields, as in eq. (\ref{Rcommutator2}), 
we obtain  
\beqa
 R_1(q,Q^2) &\,=\,& \frac{N_f N_c}{8 \pi^2} Q^2\, \left[ c_1+d_1 \left[  1 - 2\gamma \, - 2 \ln{(\frac{M}{2\pi T})}\right]   \right]  \label{dictionary1}  \\
R_2(q,Q^2)&\,=\,& \frac{N_f N_c}{8 \pi^2} \,
 \frac{Q^4}{q^2} \left[ c_1+d_1-a_0 + (1-d_1) \left[ 2 \ln{(\frac{M}{2\pi T})}  + 2\gamma \right]  \right] \,. \label{dictionary2}
\eeqa

We end this section writing the near boundary expansion for the longitudinal and transversal wave functions 
\beqa
\psi (x,u)&\,=\,& \bar q^2 \left [  A^{(0)}_t + \frac{\bar w}{ \bar q} A^{(0)}_3 \right] \, u^{1/2} \, \Big \{ a_0 \, + \, \ln{u} \,+\,   (a_1 - \frac{m^2}{2} a_0) u \,+\, ( b_1 - \frac{m^2}{2}) u  \ln{u}  \nonumber \\
 &\,+\,& {\cal O}(u^2) \,+\, {\cal O}(u^2 \ln{u})  \Big \} \label{expansionpsi} \\
\Phi_i(x,u) &\,=\, & A^{(0)}_i \left \{ 1 \,+\, ( \bar Q^2  c_1  - \frac{m^2}{4} ) u \,+\, \bar Q^2 d_1 u \ln{u}  \,+\, {\cal O}(u^2) \,+\, {\cal O}(u^2 \ln{u})  \right\}  \label{expansionphi}
\eeqa
that are obtained from eqs. (\ref{aAi}). These expansions will be useful in the analysis of the next section.


\section{ Structure functions}

In the case of an ${\cal R}$ current there is a transition between hadronic and partonic behavior, when going from low to high energies, as discussed in section {\bf II}. The partonic behaviour is characterized by non vanishing structure functions. This transition corresponds also to a change in the shape of the longitudinal and transversal potentials. At low energies, both potentials present barriers that prevent wave propagation 
from the boundary to the horizon. When increasing the energy the longitudinal potential barrier disappears while the transversal potential gets squeezed towards the boundary so that the wave propagates into the black hole and is absorbed by the horizon. 

For the flavour current, the scenario is substantially different because the D7 brane geometry changes drastically with temperature.  For temperatures lower than $0.766 M$ the D7 branes, where the gauge field lives, do not touch the horizon 
(Minkowski embedding) while for temperatures higher than  $0.766 M$ they have an induced horizon (black hole embedding). 
We now analyse the effective potentials in these two cases.  

\subsection{Low temperatures: Minkowski embedding }

\begin{figure}[ht]
\includegraphics[width=60mm]{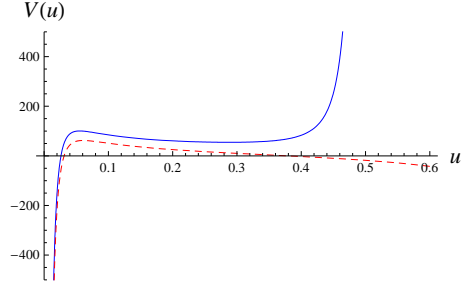}
  \caption[]{Low temperature and low energy ($\bar q\,=\,0.3 \bar Q^3$)  
  longitudinal potential of a flavour current (solid line) with $u_0\,=\,0.5$, 
  compared to that of an ${\cal R}$ current (dashed line).}
  \label{MinkD7longpot1}
\end{figure}

\begin{figure}[ht]
\includegraphics[width=60mm]{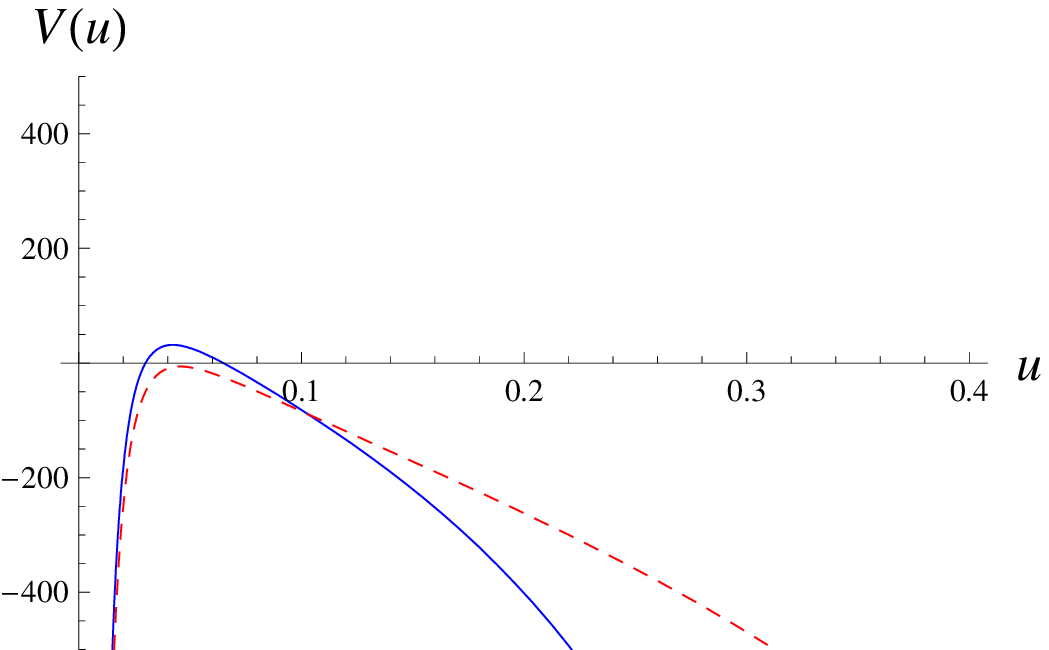} 
\hskip 1cm 
\includegraphics[width=60mm]{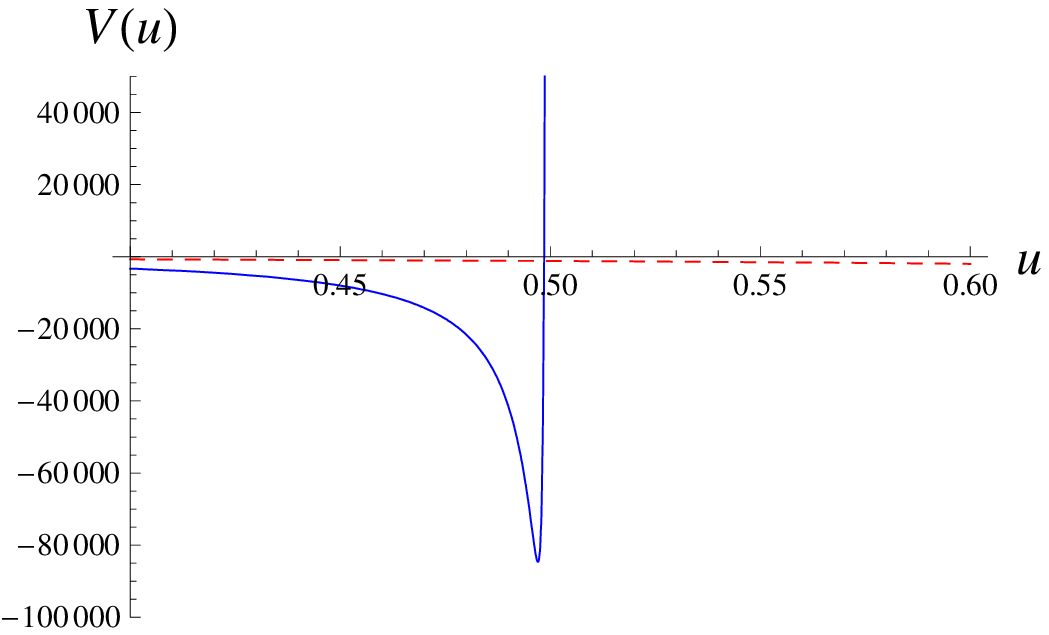}
  \caption[]{ Low temperature and high energy ($\bar q\,=\,1.6 \bar Q^3$)  longitudinal potential of a flavour current (solid line) with $u_0\,=\,0.5$, compared to that of an ${\cal R}$ current (dashed line). The left (right) panel corresponds to small (large) $u$.}
  \label{MinkD7longpot21}
\end{figure}

\begin{figure}[ht]
\includegraphics[width=60mm]{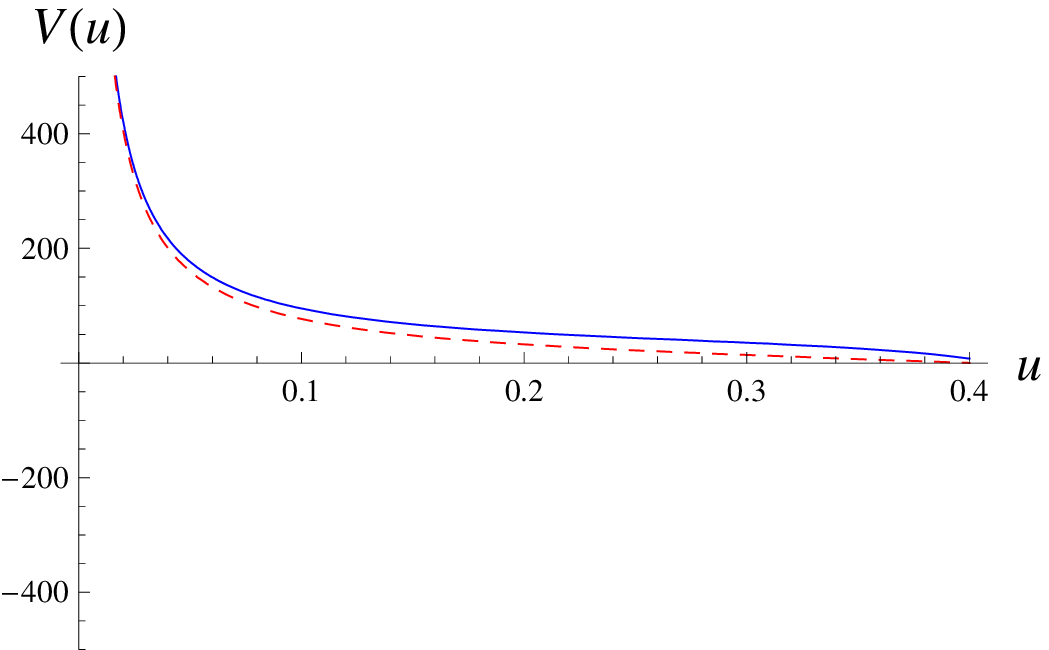}
\hskip 1cm
\includegraphics[width=60mm]{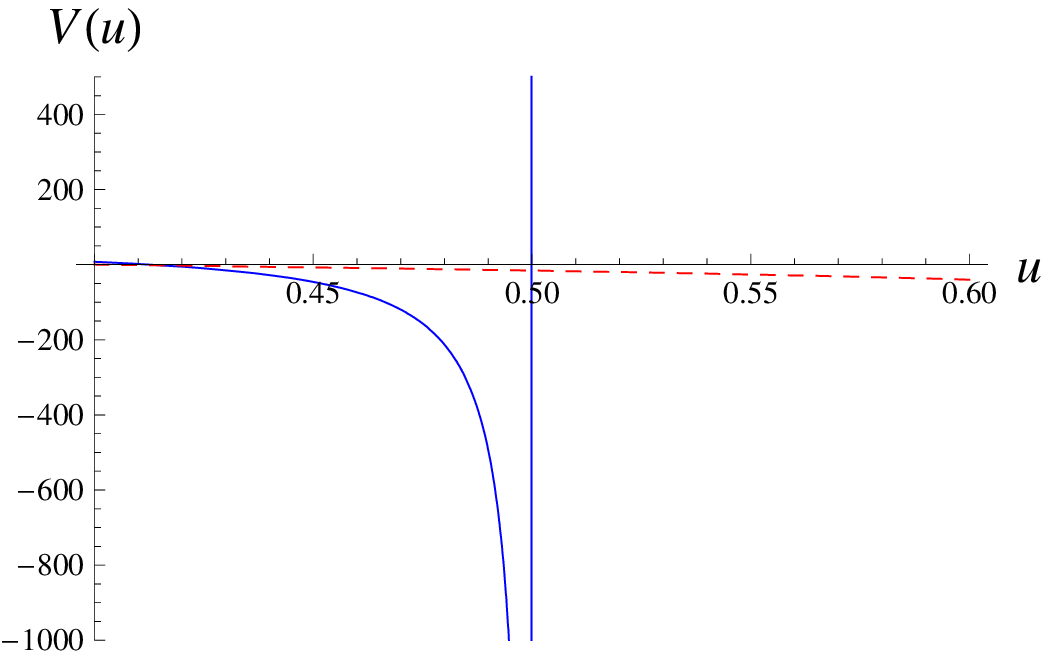}
  \caption[]{Low temperature and low energy ($\bar q\,=\,0.3 \bar Q^3$)  transversal potential of a flavour current (solid line) with $u_0\,=\,0.5$, compared to that of an ${\cal R}$ current (dashed line).  The left (right) panel corresponds to small (large) $u$.}
  \label{MinkD7transvpot11}
\end{figure}

\begin{figure}[ht]
\includegraphics[width=60mm]{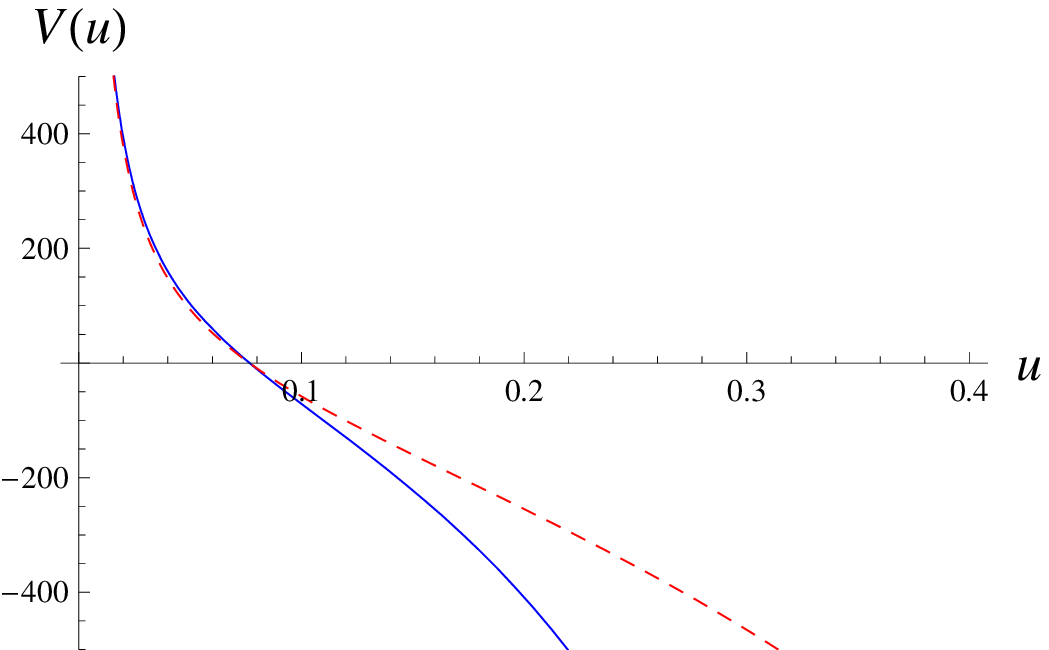}
\hskip 1cm 
\includegraphics[width=60mm]{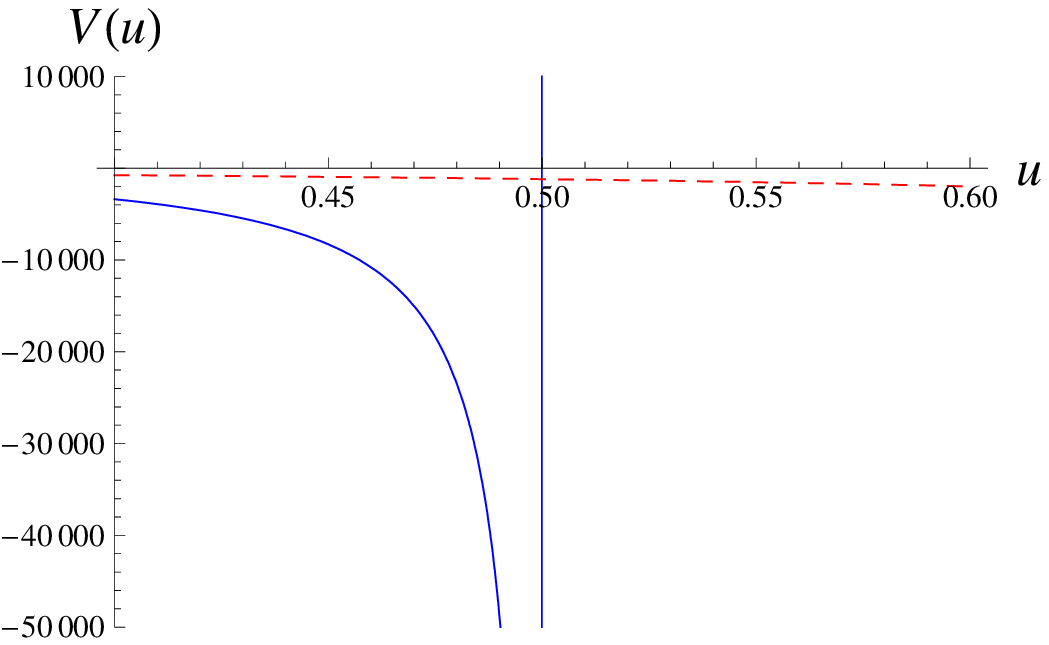}
  \caption[]{ Low temperature and high energy ($\bar q\,=\,1.6 \bar Q^3$)  transversal potential of a flavour current (solid line) with $u_0\,=\,0.5$, compared to that of an ${\cal R}$ current (dashed line).  The left (right) panel corresponds to small (large) $u$.}
  \label{MinkD7transvpot21}
\end{figure}

 At temperatures lower than $ T_c = 0.766 M$, the thermodynamically preferred D7 brane solution is the 
Minkowski embedding.  In this embedding, the radial coordinate $u$  never touches the  black hole horizon. This coordinate has a maximum value $u_0$ where the {\it tip} of the D7 branes is localized. This geometric constraint has to be present when studying the dynamics of  gauge field fluctuations. In fact, the ending condition of the radial coordinate emerges in the potential. This is illustrated in figures \ref{MinkD7longpot1} - \ref{MinkD7transvpot21} where we plot  longitudinal and transversal flavour current potentials for different energy regimes compared with the corresponding ${\cal R}$ current potentials. One can see that near the boundary the flavour current potentials behave in a similar way as the ${\cal R}$ current potentials. However, near the tip of the branes the flavour current potentials present an infinite barrier forbidding the wave to reach the black hole horizon. In contrast, the ${\cal R}$ current potentials go smoothly to the horizon.  

In the Minkowski embedding, the branes do not touch the horizon. Therefore, one can not impose {\it ingoing} boundary conditions for the wave functions. This can be seen by considering the asymptotic expansion of the longitudinal and transversal fields 
\beqa
a(x,u) &=& \left(1-\frac{u}{u_0} \right)^{\alpha} \left [ \tilde a_0 + \tilde a_1 \left(1-\frac{u}{u_0} \right) + \tilde a_2 \left(1-\frac{u}{u_0} \right)^2 + \dots \right ] \cr
A_i(x,u) &=& \left(1-\frac{u}{u_0} \right)^{\beta} \left [ \tilde A_{i,0} + \tilde A_{i,1} \left(1-\frac{u}{u_0} \right) + \tilde A_{i,2} \left(1-\frac{u}{u_0} \right)^2 + \dots \right ] \,.
\eeqa

Using these expansions in eqs. (\ref{longitudinalfieldeq}) and (\ref{transversefieldeq}) we obtain the conditions 
\beq
\alpha (\alpha + 2) = 0 \quad , \quad \beta (\beta + 1) = 0 \, . 
\eeq
with real solutions $\alpha=(0,-2)$ and $\beta=(0,-1)$. Then, the longitudinal and transversal gauge fields behave as 
{\it real} functions near the tip of the D7 branes, as well as on the boundary. So, these solutions are not compatible with an  ingoing condition. Instead,  we impose {\it regularity} of these solutions at the tip of the branes, that corresponds to  $\alpha=0$ and $\beta=0$.

This way, the infinite barrier inhibits the gravitational wave to reach the horizon. 
The wave is not absorbed but reflected at the tip of the branes, leading to stationary {\it real} solutions. 
As a consequence, the  coefficients arising in the near boundary expansion of the 
longitudinal and transversal wave functions , eqs. (\ref{expansionpsi}) and 
(\ref{expansionphi}), are also real, as well as $R_1$ and $R_2$, given by eqs. 
(\ref{dictionary1}) and (\ref{dictionary2}).  
This translates into the  vanishing of the quark contribution to the structure 
functions $F_1$ and $F_2$.
Since the infinite barriers in the effective potentials exist at any energy scale,  
we conclude that the strongly coupled plasma does not show partonic behaviour when 
interacting with a flavour current at temperatures below $T_c$. 
This result contrasts with the case of an ${\cal R}$ current at high energy, 
that indicates a partonic behaviour (associated with gluons) at any temperature.


\subsection{High temperatures: black hole embedding }

At temperatures above $T_c = 0.766 M$, the thermodynamically dominant solution is the black hole D7 embedding where 
the D7 branes touch the horizon. In this case, the gauge field waves can be absorbed by the black hole horizon, depending on the shape of the effective potentials. 
These potentials vary with the momentum scales $\bar q\,$ and $ \bar Q$ and also with the D7 brane parameter 
$m$ (mass scale). 

\begin{figure}[ht]
\includegraphics[width=60mm]{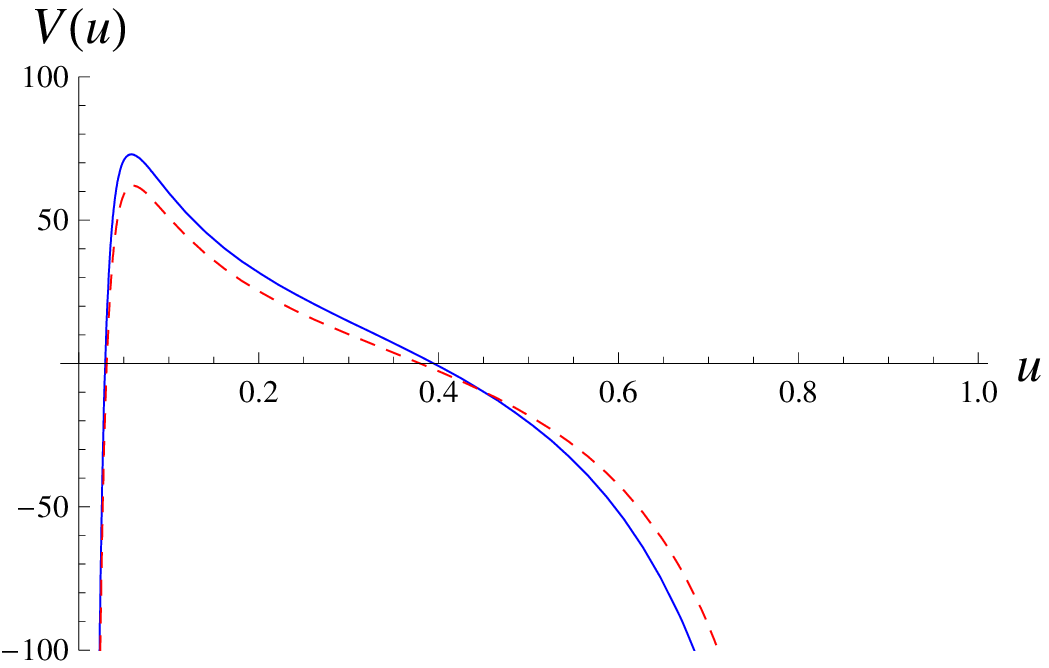}
\includegraphics[width=60mm]{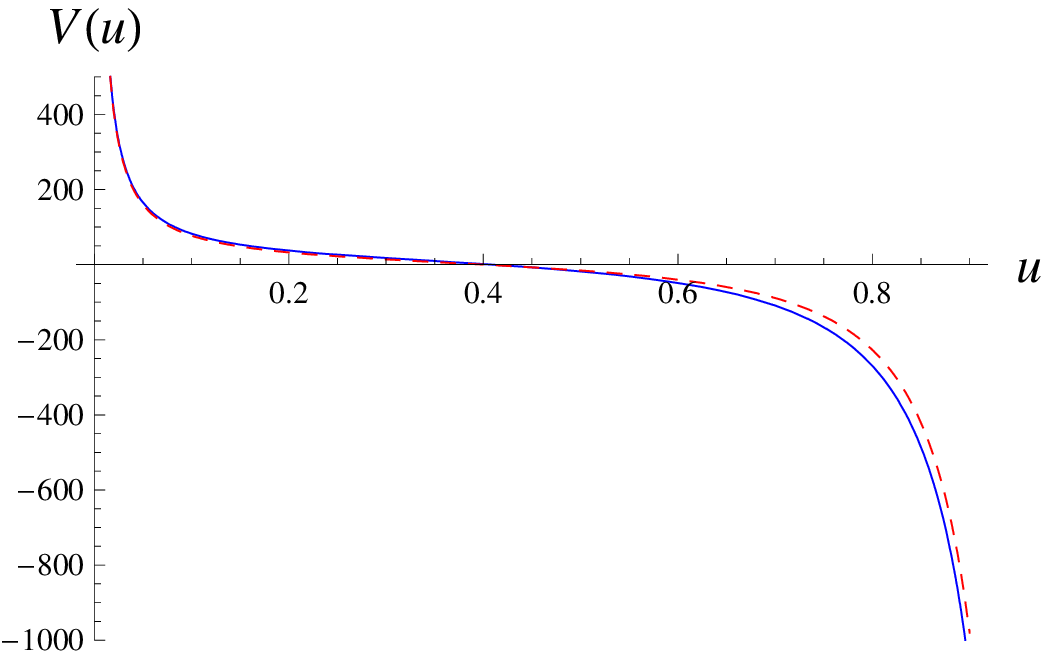}
 \caption []{High temperature and low energy ($\bar q\,=\,0.3 \bar Q^3$)  longitudinal (left panel) and
transversal (right panel) potentials of a flavour current (solid line) with $\theta_0\,=\,0.75$, compared to that of an ${\cal R}$ current (dashed line).}
  \label{BHD7longpot1}
\end{figure}

The form of the potentials is determined essentially by the ratio      
${\bar q}/{\bar Q}^3$. In the low energy case $\bar q\,\ll \, \bar Q^3$ the potentials present large barriers that inhibit propagation towards the horizon. This is illustrated in figure \ref{BHD7longpot1}, including a comparison with the ${\cal R}$ current potentials. Thus there is no relevant absorption of the flavour current, implying that the plasma structure functions vanish at low energies. 

\begin{figure}[ht]
\includegraphics[width=60mm]{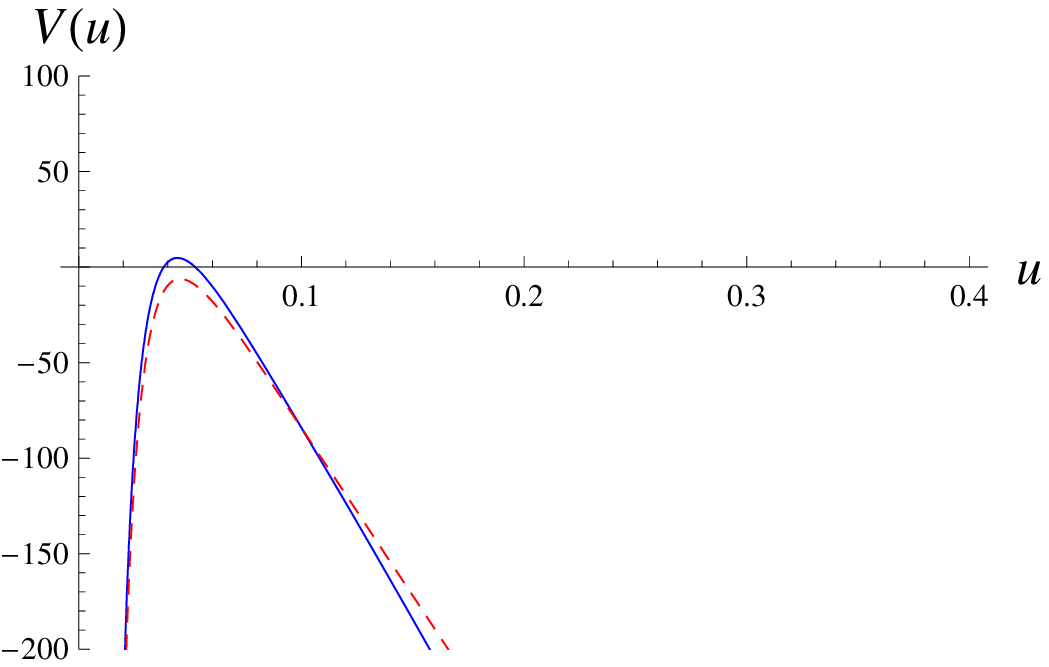}
\includegraphics[width=60mm]{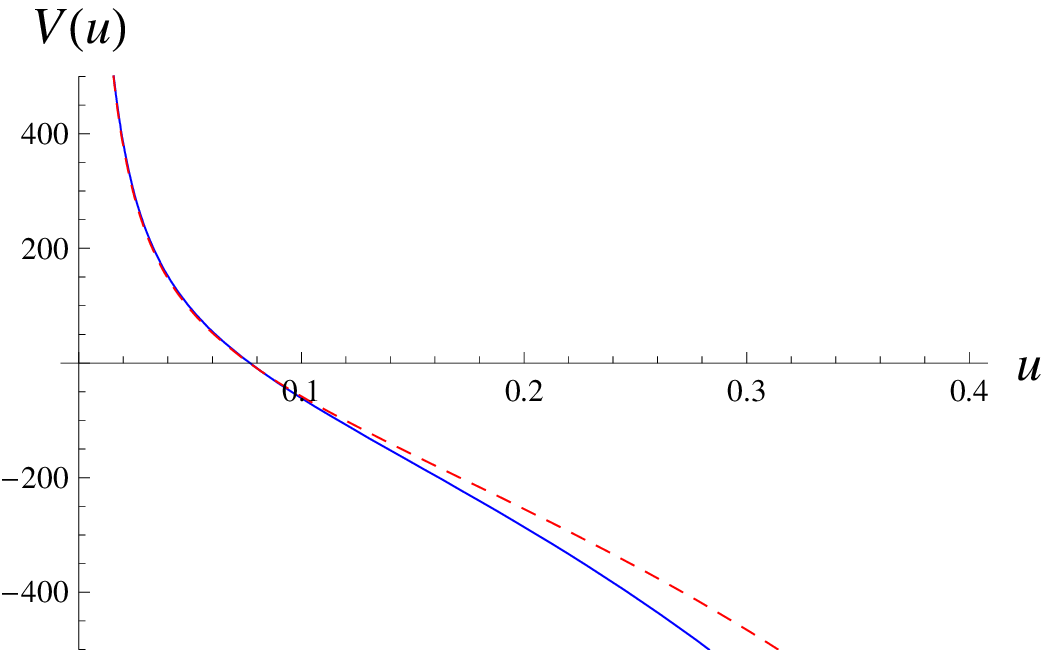}
  \caption[]{High temperature and high energy ($\bar q\,=\,1.6 \bar Q^3$)  longitudinal (left panel) and
transversal (right panel) potentials of a flavour current (solid line) with $\theta_0\,=\,0.75$, compared to that of an ${\cal R}$ current (dashed line).}
  \label{BHD7longpot2}
\end{figure}

For high energies $\bar q\,\gg \, \bar Q^3$ the longitudinal potential barrier disappears and  the transversal potential barrier gets squeezed near the boundary, as illustrated in figure \ref{BHD7longpot2}, including a comparison with the ${\cal R}$ current case.
This allows the wave absorption by the black hole horizon that can be interpreted as the absorption of the
flavour current by the quark constituents of the plasma. We will now calculate the quark contribution
to the plasma structure functions in this high energy regime.

In the black hole embedding, an analysis of the eqs. (\ref{longitudinalfieldeq}) and (\ref{transversefieldeq}) 
near the horizon $ u = 1 $ leads to 
\beqa
a(x,u) &=& (1-u)^{\pm \,\frac{i}{2} \bar w } \left[ \, \tilde a_0 + \tilde a_1 (1-u) + \tilde a_2 (1-u)^2 + \dots \,\, \right] \, , \cr 
A_i(x,u) &=&  (1-u)^{\pm \,\frac{i}{2} \bar w } \left[ \, \tilde A_{i,0} + \tilde A_{i,1} (1-u) + \tilde A_{i,2} (1-u)^2 + \dots \,\, \right] \,\,. \label{asymptgaugeBH}
\eeqa

So it is possible to impose an {\it ingoing} wave condition at the horizon, corresponding to choosing the  exponent $-i \bar w /2$.

\subsubsection{Longitudinal contribution}

Using the high energy approximation for the longitudinal potential near the boundary given in eq. (\ref{approxlongpothighenergy}) 
we have 
\beq
\partial_u^2 \psi  \,+\, \left (  \frac{1}{4u^2} \,+\, \bar q^2 u  \right ) \psi \,=\, - \frac{m^2 }{2}{ \bar q}^2 u^2 \psi \,. \label{eqpsihighenergy}
\eeq

In the black hole phase the parameter $ m $ satisfies the condition  $ m < 1.31\,$. Then, near the boundary  $m^2 u $ is very small and the term on the rhs of eq. (\ref{eqpsihighenergy}) can be neglected. Then  we obtain the solution 
\beq
\psi(x,u) \,=\,\bar q^2 \left [  A^{(0)}_t + \frac{\bar w}{ \bar q} A^{(0)}_3 \right] u^{1/2} \left [ {\cal C}_1 J_0 ( \frac23 \bar q u^{3/2}) \,+\, {\cal C}_2 Y_0 ( \frac23 \bar q u^{3/2}) \right] \, , 
\eeq

\noindent where $J_0(z)$ and $Y_0(z)$ are Bessel functions, and  ${\cal C}_1$,  ${\cal C}_2$ are complex constants. 
Analysing the behaviour of this solution for large arguments of the Bessel functions we see that a progressive wave going from the boundary to the horizon can be obtained for ${\cal C}_1\,=\,- i \, {\cal C}_2$. So, we impose this condition and assume that the solution will remain an ingoing wave at the horizon. 

In the limit $u \to 0$, relevant for the calculation of the retarded Green's function, we can expand the Bessel functions  
 finding   
\beqa
\psi(x,u) &\,=\,&\bar q^2 \left [  A^{(0)}_t + \frac{\bar w}{ \bar q} A^{(0)}_3 \right] u^{1/2} {\cal C}_2 
\Big \{ - i  \,+\, \frac{2}{\pi}  \left[ \ln ( \frac{\bar q}{3} u^{3/2} ) + \gamma \right]  \nonumber \\ 
&\,+\,& {\cal O} (\bar q^2 u^3) \,+\, {\cal O}(\bar q^2 u^3 ) \ln ( \frac{\bar q}{3} u^{3/2} ) \Big\}\,. 
\label{besselexpansionlong}
\eeqa

\noindent In the limit $u \to 0$, the perturbative corrections arising from the mass term of eq. 
(\ref{eqpsihighenergy}) will be of order $u^4$, and so can be neglected with respect to the terms in eq. 
(\ref{besselexpansionlong}). Therefore, the mass term does not contribute to the Green's function and 
hence to the DIS structure functions at high energies. 
Comparing the expansion (\ref{besselexpansionlong}) with (\ref{expansionpsi}) we find that 
\beq
{\cal C}_2 \,=\ \frac{\pi}{3} \quad \,; \quad a_0\,=\, - i \,\frac{\pi}{3} 
\,+\, \frac23 \left ( \ln{ \frac{\bar q}{3}} + \gamma \right )\,.
\eeq

The imaginary part of this expression will contribute to the DIS structure function.

\subsubsection{Transversal contribution}

In the transversal case at high energies, using (\ref{approxtransvpothighenergy}) we have 
\beq
\partial_u^2 \Phi_i \,+\, \bar q^2 u \Phi_i \,=\, - \frac{m^2}{2} \bar q^2 u^2 \Phi_i \, . \label{eqtransveraprox}
\eeq

As in the longitudinal case, we can neglect the mass term finding 
\beq
\Phi_i(x,u)\,=\, A_i^{(0)} u^{1/2} \, \left [ {\cal D}_1 J_{1/3} ( \frac23 \bar q u^{3/2}) 
\,+\, {\cal D}_2 Y_{1/3} ( \frac23 \bar q u^{3/2}) \right] \, ,
\eeq

\noindent The ingoing condition leads to ${\cal D}_1\,=\, -i {\cal D}_2 $. In the limit $ u \to 0$ we expand the Bessel functions as  
\beq
\Phi_i(x,u) = A_i^{(0)}\, {\cal D}_2 \left \{ -  \frac{\Gamma(\frac 13)}{\pi}  \left( \frac{ \bar q}{3} \right)^{-\frac 13}  
+ \frac{\sqrt{3}}{\Gamma(\frac 13)}\left[ - \sqrt{3} i  +  1 \right]   
\left( \frac{ \bar q}{3} \right)^{\frac 13}  u  + {\cal O}(\bar q^2 u^3)  \right \}\,. \label{besselexpansiontransv}
\eeq

The perturbative correction to the function $\Phi_i$ coming from the mass term is real and of order $m^2 u$. So
it can be neglected in the high energy limit. 
Now, comparing the expansion (\ref{besselexpansiontransv}) with (\ref{expansionphi})  we find 
\beqa
{\cal D}_2 =  - \frac{ \pi}{\Gamma(1/3)} \left( \frac{ \bar q}{3} \right)^{1/3} \,\, ,  \,\, 
c_1 =  - \frac{ \sqrt{3} \pi}{\Gamma^2(1/3)} \left( \frac{ \bar q}{3 {\bar Q}^3} \right)^{2/3} + i \frac{ 3 \pi}{\Gamma^2(1/3)} \left( \frac{ \bar q}{3 {\bar Q}^3} \right)^{2/3} \, , \, d_1 = 0 \, . 
\eeqa

\subsubsection{DIS structure functions} 

Using the results for the longitudinal and transversal potentials and the 
dictionary (\ref{dictionary1}),(\ref{dictionary2})  we obtain  
\beq
{R}_1 \,=\, \frac32  \, \frac{\pi}{\Gamma^2(1/3)} \, N_f N_c T^2 \left( \frac{ \bar q}{3} \right)^{2/3}\, \left[ - \frac{1}{\sqrt{3}} \,+\, i \right]  \quad , \quad {R}_2 \,=\, \frac{Q^2}{q^2} {R}_1 \, .
\eeq

Note that the imaginary part of  $a_0$ was neglected with respect to 
the imaginary part of $c_1$. Then, the quark contributions to the plasma structure functions in the high energy limit are 
\beqa
F_1(x,Q^2) &\,=\,& \frac{1}{2\pi} {\rm Im} \, {R}_1 \,\approx \,  \frac{3 }{4 \Gamma^2(1/3)} \, N_f N_c T^2 \left( \frac{ Q^2}{12 \pi T^2 x} \right)^{2/3} \nonumber \\
F_2 &\,=\,& - \frac {n \cdot q }{2 \pi T}\,  {\rm Im} \,{R}_2 \,\approx \, 2 x F_1 \,,
\eeqa

\noindent where $x$ is the Bjorken variable that in the plasma rest frame takes the form $ x = Q^2 / 2 w T\,$.

These structure functions have the same dependence in the kinematic  variables as the ${\cal R} $ current case discussed in section {\bf II}. 
However the $N_c N_f$ factor contrasts with the $ N^2_c $ factor of the ${\cal R} $ current.
Explicitly, the plasma structure functions at high temperatures arising from the flavour current and ${\cal R}$ current are related by
\beq
\frac{F_{1,2}\,[{\rm flavour}]}{F_{1,2}\,[{\cal R}]} \,\approx \, 4 \, \frac{ N_f }{N_c} \,.
\eeq

This result can be interpreted in the following way.  
The ${\cal R} $ current interacts with fields in the adjoint representation of the  $SU(N_c) $  
gauge group, that carry a number of degrees of freedom proportional to $ N^2_c $.
These fields describe gluons and their superpartners that are probed by the DIS of an ${\cal R}$ current.  
When we add the D7 probe branes, we include fields in the fundamental representation, that carry a number  
of degrees of freedom proportional to   $N_c N_f$. These fields describe the quarks and their superpartners
and are probed by the  DIS of a flavour current. Note that the ratio $N_f / N_c $ was already found in 
\cite{Mateos:2007yp} comparing the  electric conductivities associated with
a  flavour current and an $\cal R$ current. 

Note that our results for the structure functions $F_{1,2}$ at high energies hold for any value of the mass 
parameter $m$  lower than $1.31 $, which corresponds to temperatures higher than $T_c= 0.766 M$. 
When the temperature decreases to $T=T_c$, the D7 brane embedding experiments a first order transition that 
takes it from a black hole to a Minkowski phase. 
As a consequence, the imaginary part of the gauge fields arising from the ingoing wave condition disappears
leading to the vanishing of the quark contribution to the plasma structure functions, as we discussed in 
part A of this section.

\section{Conclusions} 

We studied the DIS off a $ {\cal N}= 2\,\, $ super Yang Mills plasma with flavour degrees of freedom using 
the holographic D3-D7 brane model at finite temperature. At high temperatures we found that the flavour 
current probes the partonic structure associated with the quark degrees of freedom in the same way as the 
${\cal R}$ current probes the partonic structure associated with the gluonic degrees of freedom. 
At low temperatures the flavour current is not absorbed by the quark constituents of the plasma.
As discussed above, this is a consequence of the abrupt change on the gauge fields behaviour 
far from the boundary due to the phase transition suffered by the D7 brane embedding. 

Our results contrast with the gluonic sector, probed by an ${\cal R}$ current, that shows a partonic structure, 
associated with gluons, at any temperature. This might be related to the fact that in this model there 
is an effective infrared mass scale for the quark sector, but not for the gluon sector. 

As we explained in this article, the non partonic absorption of the flavour current at low temperatures 
is a consequence of the absence of an imaginary part in the retarded Green's function. It is important 
to remark that a real Green's function may have poles and then give rise to imaginary contributions in 
the form of delta peaks. This kind of contribution was analysed in ref. \cite{Myers:2007we} for a 
time-like flavour current in the low temperature Minkowski embedding. In that reference the delta peaks 
of the spectral function where interpreted in terms of the meson spectra. Here we have worked with a 
space-like flavour current and we did not consider this  kind of contribution, since we were only 
interested in the partonic structure of the plasma. 

After the first version of this article, ref. \cite{Iancu:2009py} proposed a different mechanism for the absorption of a flavour current at low temperatures, based on vector meson production. This process contributes to the $ {\cal N}= 2\,\, $ super Yang Mills plasma structure functions of the D3-D7 brane model.

We worked in the probe approximation of the D3-D7 brane model in which the backreaction 
of the D7 branes is not taken into account. A review of the D3-D7 brane model beyond the probe limit was done in  
\cite{Erdmenger:2007cm}. 
A recent proposal to include backreaction in the D3-D7 quark-gluon plasma appeared in \cite{Bigazzi:2009bk}. 
It might be interesting to investigate the effect of these backreaction corrections on the plasma structure 
functions and on its partonic behaviour.


\bigskip

\noindent {\bf Acknowledgments:} This work was originated in discussions with  Edmond Iancu and  Al Mueller.
We also thank them for reading the manuscript and making various interesting comments.   
The authors are partially supported by CAPES, CNPq and FAPERJ.

 \end{document}